\documentclass[9pt, twocolumn,superscriptaddress]{revtex4}

\usepackage{graphicx}
\usepackage{soul}
\usepackage{amsmath}
\usepackage{bm}
\usepackage{siunitx}
\usepackage{makecell}
\usepackage{hyperref}
\usepackage[capitalize]{cleveref}

\begin{document}

\title{A Bayesian perspective on single-shot laser characterization}


\author{J. Esslinger}
\thanks{These authors contributed equally.}
\affiliation{Ludwig--Maximilians--Universität München, Am Coulombwall 1, 85748 Garching, Germany}
\author{N. Weiße}
\thanks{These authors contributed equally.}
\affiliation{Ludwig--Maximilians--Universität München, Am Coulombwall 1, 85748 Garching, Germany}
\author{C. Eberle}
\affiliation{Ludwig--Maximilians--Universität München, Am Coulombwall 1, 85748 Garching, Germany}
\author{J. Schröder}
\affiliation{Ludwig--Maximilians--Universität München, Am Coulombwall 1, 85748 Garching, Germany}
\author{S. Howard}
\affiliation{Ludwig--Maximilians--Universität München, Am Coulombwall 1, 85748 Garching, Germany}
\affiliation{Department of Physics, Clarendon Laboratory, University of Oxford, Parks Road, Oxford OX1 3PU, United Kingdom}
\author{P. Norreys}
\affiliation{Department of Physics, Clarendon Laboratory, University of Oxford, Parks Road, Oxford OX1 3PU, United Kingdom}
\author{S. Karsch}
\affiliation{Ludwig--Maximilians--Universität München, Am Coulombwall 1, 85748 Garching, Germany}
\author{A. Döpp}
\email[]{a.doepp@lmu.de}
\affiliation{Ludwig--Maximilians--Universität München, Am Coulombwall 1, 85748 Garching, Germany}
\affiliation{Department of Physics, Clarendon Laboratory, University of Oxford, Parks Road, Oxford OX1 3PU, United Kingdom}


\begin{abstract}
We introduce a Bayesian framework for measuring spatio-temporal couplings (STCs) in ultra-intense lasers that reconceptualizes what constitutes a 'single-shot' measurement. Moving beyond traditional distinctions between single- and multi-shot devices, our approach provides rigorous criteria for determining when measurements can truly resolve individual laser shots rather than statistical averages. This framework shows that single-shot capability is not an intrinsic device property but emerges from the relationship between measurement precision and inherent parameter variability. Implementing this approach with a new measurement device at the ATLAS-3000 petawatt laser, we provide the first quantitative uncertainty bounds on pulse front tilt and curvature. Notably, we observe that our Bayesian method reduces uncertainty by up to 60\% compared to traditional approaches. Through this analysis, we reveal how the interplay between measurement precision and intrinsic system variability defines achievable resolution—insights that have direct implications for applications where precise control of laser-matter interaction is critical.
\end{abstract}

\maketitle

\section{Introduction}
The characterization of complex physical systems presents a fundamental challenge in metrology: how to capture high-dimensional information using inherently limited measurement devices. This challenge is exemplified in ultra-intense lasers, where petawatt-scale powers and focused intensities exceeding $10^{23}$ W/cm$^2$ push the boundaries of measurement capabilities \cite{Danson.2015,yoon2021realization}. To fully characterize such lasers requires measuring the electromagnetic vector field across three spatial dimensions and time, yet most detectors are confined to two-dimensional views with temporal resolution far below the pulse duration.
Traditional approaches using sequential scanning techniques \cite{bowlan07,bowlan08,gallet14-2,alonso10,alonso12-1,alonso12-2,alonso13,miranda14,pariente16,borot2018spatio} are strictly only suitable when laser systems exhibit neither drifts nor significant shot-to-shot fluctuations. When these conditions are not met, 'few-shot' methods like FALCON \cite{weisse2023b} and IMPALA \cite{smartsev24} trade resolution for acquisition time. For true instantaneous characterization, the field has pursued 'single-shot' measurements, though this ideal often proves elusive and comes with significant trade-offs \cite{gabolde06,dorrer18,kim21,tang2022,howard23}. Crucially, all current methods lack uncertainty quantification \cite{alonso2024space}—particularly important as measurements are typically performed near the noise floor of intrinsic fluctuations.

These challenges prompt us to reconsider laser characterization through the lens of Bayesian inference. For an isolated pulse absent any prior knowledge of its properties, capturing all information in a single shot offers clear advantages. However, many real-world scenarios involve contextual knowledge about the laser, be it from system specifications, previous measurements, or characteristic drift patterns. This is why, in our framework, each measurement represents an update to our understanding of the system, building upon prior information rather than starting from scratch. 

Consider FALCON, which measures wavefronts sequentially at different wavelengths: is it 'single-shot' because it takes an image every shot, or 'multi-shot' because it needs multiple measurements for full spectral coverage? After how many filters is a measurement "complete"? Such questions have no clear answer in traditional frameworks. Rather than attempting to capture all information in one shot, we recognize that laser systems typically exhibit dynamics at separated time scales: slow drifts over minutes to hours, and rapid fluctuations above the laser's repetition rate. While slow evolution can be predicted using time series models \cite{dopp2023data}, high-frequency components manifest as stochastic shot-to-shot variations requiring instantaneous measurement. This distinction provides rigorous criteria for determining when measurements can truly resolve individual laser states rather than statistical averages. Understanding these shot-to-shot variations is crucial, especially for applications that make use of tailored STCs, ranging from particle acceleration \cite{debus19,palastro20,caizergues20} to attosecond pulse generation \cite{vincenti12,quere14}.

In this paper, we present a rigorous mathematical framework demonstrating how 'single-shot' capability emerges from the relationship between measurement precision and intrinsic system variability. \cref{sec:theory} provides analytical tools for studying measurement system behavior and parameter estimation, while \cref{sec:experiment} showcases a demonstration for spatio-temporal coupling measurements in ultra-intense lasers. To our knowledge, this represents the first proper probabilistic treatment of single-shot laser characterization.
\section{Theory}\label{sec:theory}

As mentioned earlier, we describe the evolution of any measured parameter as a combination of deterministic and stochastic components. More formally, the state of any parameter $x$ at times $t_k$ can be modeled as

\begin{equation}
    x_k = f(t_k) + \varepsilon_k \label{eq:decomposition}
\end{equation}

where $f(t_k)$ represents predictable changes in the system and $\varepsilon_k \sim \mathcal{N}(0,\sigma^2_\text{stoch})$ are independent, normally distributed random variables characterizing shot-to-shot fluctuations. The normal distribution naturally arises here as the aggregate effect of many small independent fluctuations tends to be normally distributed due to the Central Limit Theorem. The variance parameter $\sigma_\text{stoch}^2$ characterizes the intrinsic stochasticity of the system, that is, fluctuations that cannot be predicted ahead.

Each measurement $y_k$ comes with its own uncertainty,

\begin{equation}
    y_k = x_k + \epsilon_k =  f(t_k) + \varepsilon_k + \epsilon_k 
\end{equation}

where $\epsilon_k \sim \mathcal{N}(0,\sigma^2_\text{meas})$ is the measurement noise. This measurement uncertainty, combined with the intrinsic stochasticity, determines our ability to resolve changes in the system state. Understanding this interplay requires a Bayesian perspective, where each measurement updates our knowledge of both the deterministic and stochastic components.

\begin{figure}[t]
    \centering
    \includegraphics[width=\columnwidth]{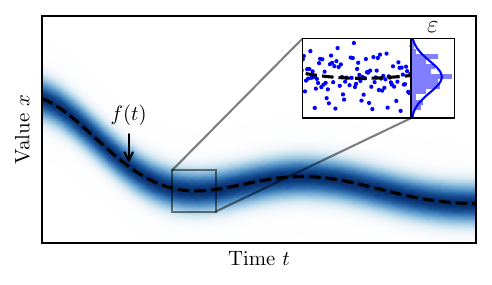}
    \caption{Decomposition of measurements into deterministic and stochastic components. The dashed black line shows the predictable trend $f(t)$, while the blue density represents the measurement spread due to intrinsic stochasticity. The inset shows a zoomed region with individual measurements (blue dots) and their statistical distribution (histogram), demonstrating how the stochastic component $\varepsilon_k \sim \mathcal{N}(0,\sigma^2_\text{stoch})$ manifests in the measurements.}
    \label{fig:decomposition}
\end{figure}

\subsection{Bayesian state updates}\label{sec:Bayes}

At the heart of Bayesian inference lies Bayes' theorem, which provides a formal method for updating our beliefs about a state variable $x_k$ given new measurement data $y_k$:
\begin{equation}
p(x_k|y_k) = \frac{p(y_k|x_k)p(x_k)}{p(y_k)}
\label{eq:bayes_theorem}
\end{equation}
Here, $p(x_k|y_k)$ is the posterior probability (our updated belief), $p(y_k|x_k)$ is the likelihood (the probability of observing the measurement given the state), $p(x_k)$ is the prior probability (our initial belief), and $p(y_k)$ is the evidence (a normalization factor). We make the recursive Markov assumption for the prior, i.e. that each new state depends only on the immediately previous state, not the entire history of states. The evidence can be expressed as an integral over all possible values of the state
\begin{equation}
p(y_k) = \int p(y_k|x_k)p(x_k) dx_k.
\label{eq:evidence_integral}
\end{equation}
In practice, solving this integral can be challenging.
But given a Gaussian model for both the stochastic component and measurement uncertainty, we can solve \cref{eq:bayes_theorem} analytically due to the self-conjugacy of Gaussian distributions. For Gaussian distributions, with prior $p(x_k) = \mathcal{N}(\mu_{k|k-1}, \sigma^2_{k|k-1})$ and likelihood $p(y_k|x_k) = \mathcal{N}(x_k, \sigma^2_\text{meas})$, the posterior is proportional to:
$$p(x_k|y_k) \propto \exp\left(-\frac{(x_k-\mu_{k|k-1})^2}{2\sigma^2_{k|k-1}} - \frac{(x_k-y_k)^2}{2\sigma^2_\text{meas}}\right)$$
Here, $\mu_{k|k-1}$ and $\sigma^2_{k|k-1}$ represent our prediction (prior) estimate and uncertainty before the k-th measurement, while $y_k$ is the observed measurement, and $\sigma^2_\text{meas}$ represents the measurement uncertainty.
Completing the square in the exponent shows that the mean and variance of this posterior Gaussian distribution are given by:
\begin{equation}
\mu_{k|k} = (1-\gamma_k)\cdot\mu_{k|k-1} + \gamma_k\cdot y_k
\label{eq:posterior_mean}
\end{equation}
\begin{equation}
\sigma^2_{k|k} = \left( \frac{1}{\sigma^2_{k|k-1}} + \frac{1}{\sigma^2_\text{meas}}\right)^{-1}
\label{eq:posterior_variance}
\end{equation}
Here we introduced the update weight $\gamma_k$
\begin{equation}
\gamma_k = \frac{\sigma^2_{k|k-1}}{\sigma^2_{k|k-1} + \sigma^2_\text{meas}}.
\label{eq:kalman_gain}
\end{equation}

Note that the prior mean $\mu_{k|k-1}$ and variance $\sigma^2_{k|k-1}$ are based on our prediction what to expect, see \cref{sec:prediction}. The lower limit on the prior variance is given by the unpredictable intrinsic stochasticity $\sigma_{\text{stoch}}$ of the system.

\subsection{Asymptotic Behavior and Measurement Regimes}\label{sec:steady-state}

The update weight defined in \cref{eq:kalman_gain} encapsulates several key insights. First, it quantifies how our uncertainty evolves with new measurements, accounting for both our prior knowledge and the system's inherent variability. Second, it reveals the interplay between measurement precision and intrinsic stochasticity in determining our final uncertainty.
The behavior of this equation in limit cases reveals its consistency with established principles. As $\sigma^2_{\text{prior}} \to \infty$, representing complete initial ignorance, we recover the classical result: $\sigma^2_{\text{posterior}} \approx \sigma^2_{\text{meas}}$. When $\sigma^2_{\text{stoch}} \to 0$, indicating a deterministic system, $n$ repeated measurements can reduce uncertainty indefinitely following $\sigma^2_{\text{posterior}} = \sigma^2_{\text{meas}}/n$. 

Between these cases, for $\sigma^2_{\text{stoch}} > 0$ and a sequence of measurements, the asymptotic behavior of our system reveals a stationary solution for which the posterior variance ceases to reduce further:
\begin{align}
    \sigma^2_{\infty|\infty} &=  \frac{\sqrt{1 + 4(\sigma^2_{\text{meas}}/\sigma^2_{\text{stoch}})} - 1}{2}  \cdot\sigma^2_{\text{stoch}} \nonumber \\
    &= \frac{1-\gamma}{\gamma}\cdot\sigma^2_{\text{stoch}} 
    \label{eq:posterior_variance}
\end{align}

\begin{figure}[t]
    \centering
    \includegraphics[width=\columnwidth]{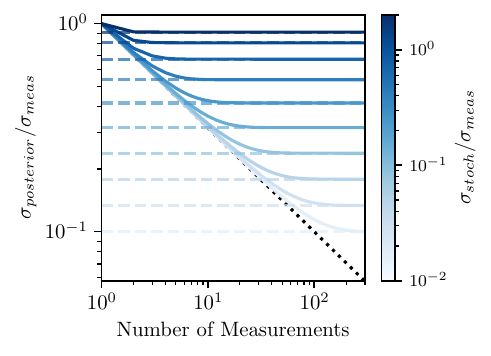}
    \caption{Evolution of the normalized posterior standard deviation ($\sigma_\text{posterior}/\sigma_\text{meas}$) over iterations for different ratios of process to measurement noise ($\sigma_\text{stoch}/\sigma_\text{meas}$, shown by color). Solid lines represent the Bayesian update process, while dashed lines indicate the corresponding asymptotic limits. The convergence rate and final value depend strongly on the noise ratio, with the dotted line indicating the ideal trend for vanishing intrinsic stochasticity.}
    \label{fig:posterior_variance_evolution}
\end{figure}

This asymptotic limit delineates a critical threshold when $\sigma^2_{\text{meas}} = 2\sigma^2_{\text{stoch}}$, distinguishing two regimes of measurement capabilities:
\begin{description}
\item[Distribution-Centric Regime] ($\sigma^2_{\text{meas}} > 2\sigma^2_{\text{stoch}}$):
In this regime, the prediction relies more on the prior and the posterior variance remains always greater than the inherent system variability. For $\sigma^2_{\text{meas}} \gg \sigma^2_{\text{stoch}}$ the limit is approximately $\sigma^2_{\infty|\infty}\approx \sigma_{\text{stoch}}\cdot \sigma_{\text{meas}} = \sigma^2_{\text{stoch}}\cdot( \sigma_{\text{meas}}/\sigma_{\text{stoch}})$. In this regime of precision, there is no fundamental advantage of a single-shot device except for shorter measurement time, because a multi-shot device can accumulate equivalent information about the average properties of the laser over time.
\item[State-Resolving Regime] ($\sigma^2_{\text{meas}} < 2\sigma^2_{\text{stoch}}$):
In this regime, we achieve sufficient precision to meaningfully distinguish single shots. Here a measurement can provide information about the instantaneous system state that allows our knowledge to surpass its average behavior, i.e. $\sigma^2_{\infty|\infty}<\sigma_{\text{stoch}}$. Note that the stationary solution of the update weight in this regime $\gamma = \sigma^2_{\infty|\infty}/(\sigma^2_{\infty|\infty} + \sigma^2_{\text{meas}})>0.5$, meaning during Bayesian inference the measurement weight amounts to more than 50\% and the prior contribution falls below 50\%; so, a single measurement in this regime takes less advantage of prior information.
\end{description}
This threshold helps us to better understand what truly constitutes 'single-shot' measurement devices. Rather than being an intrinsic property of the measurement apparatus, the single-shot capability is now understood as a relationship between measurement precision and intrinsic stochasticity. 
Instead of relying on arbitrary definitions of measurement completeness, we can now precisely quantify when a device can meaningfully resolve individual system states. While there is certainly a significant practical advantage in acquiring as much diverse information as possible within in a single shot, additional information only emerges as the measurement precision crosses this critical threshold relative to the system's inherent variability. Our analysis also confirms the intuitive notion that a state-resolving device must inherently rely more on the data from a single measurement than on prior information. Note that this is a general statement that includes \textit{any} type of prior information, including sophisticated artificial neural network prediction.

\subsection{Relationship between Noise and Sampling Frequency}\label{sec:frequency-analysis}

A critical aspect in laser metrology is understanding the interplay between intrinsic stochasticity $\sigma^2_{\text{stoch}}$ and our measurement paradigm. What appears as stochastic noise in a laser system often represents deterministic processes occurring at timescales beyond our measurement resolution. This phenomenon can lead to a deterministic, dynamic system masquerading as a stationary, stochastic one if the sampling frequency is insufficient. This is analogous to aliasing in signal processing, where undersampling can lead to misinterpretation of high-frequency components as lower-frequency variations. To better understand this, we consider the relationship between our measurement frequency and the characteristic frequencies of the laser system's dynamics. 

\begin{figure}[t]
    \centering
    \includegraphics[width=1\columnwidth]{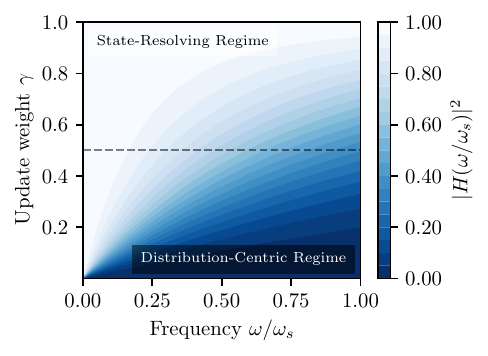}
    \caption{Frequency response $|H(\omega/\omega_s)|^2$ as a function of normalized frequency $\omega/\omega_s$ and update weight $\gamma$. This plot illustrates how the system's frequency response changes with different update weights, highlighting the trade-off between noise reduction and dynamic response in the Bayesian inference process.}
    \label{fig:magnitude_response}
\end{figure}

In its asymptotic limit, the Bayesian update \cref{eq:posterior_mean} behaves like an exponential moving average filter with the update weight $\gamma$ as time constant. The frequency response of the system is then given as
\begin{equation}
    \| H(\omega/\omega_s) \|^2 = \gamma^2 / (1-2(1-\gamma)\cos(\omega/\omega_s)+(1-\gamma)^2)
    \label{eq:magnitude-response}
\end{equation}
where $\omega_s$ is the sampling rate. From this equation we see how a smaller update weight leads to the attenuation of higher frequencies and thereby obfuscates dynamics, see \cref{fig:magnitude_response}. At our regime threshold, we find that the Nyquist frequency $\omega_s/2$ is attenuated to about 67\%. We can roughly quantify the ability to resolve frequencies using the cutoff frequency $\omega_c = \arccos((\gamma^2+2\gamma-2)/(2\gamma-2))\omega_s$ for which the spectral power decays to $\| H(\omega_c/\omega_s) \|^2 = 1/2$. Shot-to-shot dynamics at frequencies above this cutoff tend to not be resolved and will appear stochastic. 

The concept of intrinsic stochasticity, therefore, emerges as a relative construct, intricately tied to our measurement capabilities rather than an absolute property of the laser system. In practical terms, a 1 Hz laser system may for instance exhibit fluctuations at tens or hundreds of Hz due to mechanical vibrations. As the laser's repetition rate, and with it the minimum sampling rate, lies below these frequencies, these deterministic oscillations manifest as apparent randomness in our measurements. This effect propagates across timescales: as we increase reliance on prior measurements, we inevitably sacrifice frequency resolution, potentially misclassifying even slower drifts as 'noise'.

This perspective reframes our understanding of intrinsic stochasticity within the context of laser metrology as a delineation between deterministic and probabilistic modeling. This boundary is not fixed but dynamic, shaped by our measurement capabilities and analysis methods. As we enhance our ability to resolve and model faster laser dynamics, we expand the deterministic realm of our understanding. Conversely, unresolved dynamics - whether due to sampling limitations or genuinely stochastic processes - fall into the domain of intrinsic stochasticity. This perspective transforms the intrinsic stochasticity estimation from a mere quantification of uncertainty into a crucial tool for defining the limits of our predictive capabilities. It determines whether we can make precise statements about individual laser shots or must resort to statistical predictions about consecutive shots.

\subsection{Estimating Intrinsic Stochasticity}\label{sec:adaptive_process_noise}

With this refined understanding of noise in laser systems, we can now discuss how to learn to delineate the boundary between resolved dynamics and unresolved fluctuations in the laser's behavior. 

Given knowledge of the measurement noise, typically obtained through device calibration (see \cref{sec:experiment}-\ref{sec:measurement-noise}), we can estimate the intrinsic stochasticity by leveraging the discrepancy between what we predict and what we observe. This residual serves as a measure of model misspecification and forms the foundation of our adaptive estimation framework. For simplicity, we again sketch the derivation for a single variable with  \(\mu_{\text{meas}}\) and \(\sigma^2_{\text{meas}}\) as the mean and variance of the current measurement, as well as \(\mu_{\text{posterior}}\) and \(\sigma^2_{\text{posterior}}\) as the updated estimates for the parameter and its variance.

The residual \(\Delta \mu\) is the difference between the predicted state mean and the measurement mean 
\begin{equation}
    \Delta \mu = \mu_{\text{meas}} - \mu_{\text{pred}}
\end{equation}
 and the variance of the residual combines the variances of the prediction and the measurement
 \begin{equation}
     \sigma^2_{\Delta \mu} = \sigma^2_{\text{pred}} + \sigma^2_{\text{meas}} \approx \sigma^2_{\text{posterior}} + \sigma^2_{\text{stoch}} + \sigma^2_{\text{meas}} 
     \label{eq:variance_residual}
 \end{equation}
The approximation used here is motivated in \cref{sec:prediction}. From here, the likelihood of the measurement given the current model state is
\begin{equation}
   p(\Delta \mu, \sigma^2_{\Delta \mu}) = \frac{1}{\sqrt{2\pi \sigma^2_{\Delta \mu}}} \exp\left( -\frac{(\Delta \mu)^2}{2 \sigma^2_{\Delta \mu}} \right) 
   \label{eq:likelihood}
\end{equation}

The question is how to change $\sigma^2_{\text{stoch}}$ to maximize the likelihood. For this, we consider the roots of the log-likelihood 
\[ \frac{\partial \log p(\Delta \mu, \sigma^2_{\Delta \mu})}{\partial \sigma^2_{\text{stoch}}} = 0
\]
Solving this equation, while dropping the dependency of $\sigma^2_{\text{posterior}}$ on the intrinsic stochasticity itself in \cref{eq:posterior_variance} as higher order effect, we find the rather intuitive result
\[
\sigma^2_{\text{stoch}} \approx  (\Delta \mu)^2 - \sigma^2_{\text{posterior}} - \sigma^2_{\text{meas}}.
\]

However, this example only considers a single measurement, which is clearly insufficient to estimate the actual value of \(\sigma^2_{\text{stoch}} \); even for a perfect prediction of the mean and no measurement noise the residual is only a sample from the distribution \( \Delta \mu \sim \mathcal{N}(0,\sigma^2_{\text{stoch}})\). Assuming that the intrinsic stochasticity is constant or slowly varying, we can use a sequence of $n$ measurements to achieve an estimate with variance  $\sigma^2_{\text{stoch}}/n$.

It should be noted that the sample variance of the residuals can sometimes be smaller than the sum of the known variances. This can lead to physically meaningless results. Even with this, a variance matrix needs to be positive semidefinite, which might not be the case all the time with this simple approximation. Therefore, we use this analytic equation as a starting point to fit the intrinsic stochasticity by maximizing the log-likelihood numerically via Stochastic Gradient Descent. For simplicity we do not model correlations between modes and hence assume the intrinsic stochasticity matrix to be diagonal. This way, the reconstructed uncertainties can be considered upper bound values.

This estimation of intrinsic stochasticity is deeply intertwined with our ability to separate deterministic and stochastic components. The decomposition is only meaningful if we can accurately estimate the deterministic component $f(t_k)$ - otherwise, errors in estimating $f(t_k)$ will be absorbed into our estimate of $\varepsilon_k$, leading to biased estimates of $\sigma^2_\text{stoch}$. The challenge becomes particularly acute in the state-resolving regime identified in \cref{sec:steady-state}: when our measurement precision is sufficient to resolve individual states, each measurement captures not just $f(t_k)$ but also (part of) the instantaneous stochastic fluctuation $\varepsilon_k$. Using such measurements directly for prediction would erroneously treat these random fluctuations as part of the deterministic evolution. Because of this, pure state-space models like the adaptive Kalman filter \cite{mohamed1999adaptive} are not applicable to our problem; they systematically overestimate the system's noise. To address this, in the next section, we focus on estimating the deterministic component of the system evolution, before combining it with our stochasticity estimates to form a complete prediction model.

\subsection{Local Linear Approximation}\label{sec:local-linear}
For sufficiently small time intervals $\Delta t$, we can approximate the deterministic evolution locally by a linear function:
\begin{equation}
f(t_k + \Delta t) \approx x_k + v_k \Delta t
\end{equation}
where $x_k$ represents the true state and $v_k$ the instantaneous rate of change. This linear approximation is valid when changes in the deterministic component are small compared to stochastic fluctuations, i.e., $|v_k\Delta t| \ll \sigma_\text{stoch}$. Under this condition, any errors from nonlinear behavior are negligible compared to the intrinsic stochasticity we aim to characterize. Using Holt's linear exponential smoothing \cite{holt2004forecasting} with additive trends as robust framework for tracking evolving systems while filtering out noise, our best estimate $\hat{x}_k$ of this state evolves as:
\begin{equation}
\hat{x}_{k+1} = (1-\alpha) \cdot [\hat{x}_k + \hat{v}_k\Delta t] +\ \alpha \cdot \mu_{k|k}
\end{equation}
\begin{equation}
\hat{v}_{k+1} = (1-\beta) \cdot \hat{v}_k + \beta\cdot\frac{\hat{x}_{k+1}-\hat{x}_k}{\Delta t}
\end{equation}
where $\mu_{k|k}$ is the posterior mean from our measurements up to time $t_k$. The parameters $\alpha$ and $\beta$ are learned by maximizing the log-likelihood over a finite window:
\begin{equation}
\begin{aligned}
(\hat\alpha, \hat\beta) &= \underset{\alpha, \beta}{\text{argmax}} \, \sum_{k=1}^N \log p(y_k | \hat{x}_k(\alpha, \beta)) \\
&= \underset{\alpha, \beta}{\text{argmin}} \, \sum_{k=1}^N (y_k - \hat{x}_k(\alpha, \beta))^2
\end{aligned}
\end{equation}
where the equality follows from our Gaussian assumptions. By using a finite window, we assume local stationarity of the parameters while allowing the model to adapt to changes in system behavior. Note that our model assumes that variances are approximately constant, which is the case once the measurement system approaches the steady state discussed in \cref{sec:steady-state}.

More sophisticated approaches to learn $f(t)$ could be considered. Higher-order models or neural networks, for instance, could learn complex temporal patterns in the system's evolution. However, such approaches come with their own drawbacks: The more complex the model, the more additional parameters need to be estimated, potentially making the decomposition between deterministic and stochastic components less stable. This can for instance lead to confusion between genuine patterns and spurious correlations in the noise. As our experimental results in \cref{sec:results} demonstrate, the local linear model provides a pragmatic balance that can accurately separate predictable changes from stochastic variations, while its simple, recursive structure makes it compatible with real-time evaluation.

\subsection{Prediction Model}\label{sec:prediction}

Combining our local linear approximation from \cref{sec:local-linear} with the estimated stochastic component from \cref{sec:adaptive_process_noise} according to \cref{eq:decomposition} yields a probabilistic prediction model, whose mean follows the deterministic model, i.e., $\mu_{k+1|k}=\hat x_{k+1}$. The prediction's variance is calculated as the combination of the uncorrelated uncertainties of the noise and the local linear model:
\begin{equation}
\sigma^2_{k+1|k} = \sigma^2_\text{stoch} + \sigma^2_\text{pred}(\Delta t) \approx  \sigma^2_\text{stoch} + \sigma^2_{k|k}
\end{equation}
Here we assumed that the model noise is dominated by our uncertainty in the true state, given by our posterior variance, and neglected the uncertainty on the trend of the local linear model. This is justified when the time interval $\Delta t$ is sufficiently small, as the model uncertainty grows with $ (\Delta t)^2 \cdot \text{Var}(\hat{v}_k)$.

For small measurement intervals, we can safely neglect the trend's uncertainty term. This condition is typically satisfied in our application, where measurement rates are high compared to the timescale of systematic drifts. Even when this approximation introduces some error, it manifests as an apparent increase in the estimated intrinsic stochasticity, ensuring our uncertainty estimates remain conservative.

The probabilistic predictions from this combined model serve as our prior for subsequent Bayesian state updates, completing the loop back to \cref{sec:Bayes}.

\section{Experiment}\label{sec:experiment}

\begin{figure}[t]
    \centering
    \includegraphics[width=\linewidth]{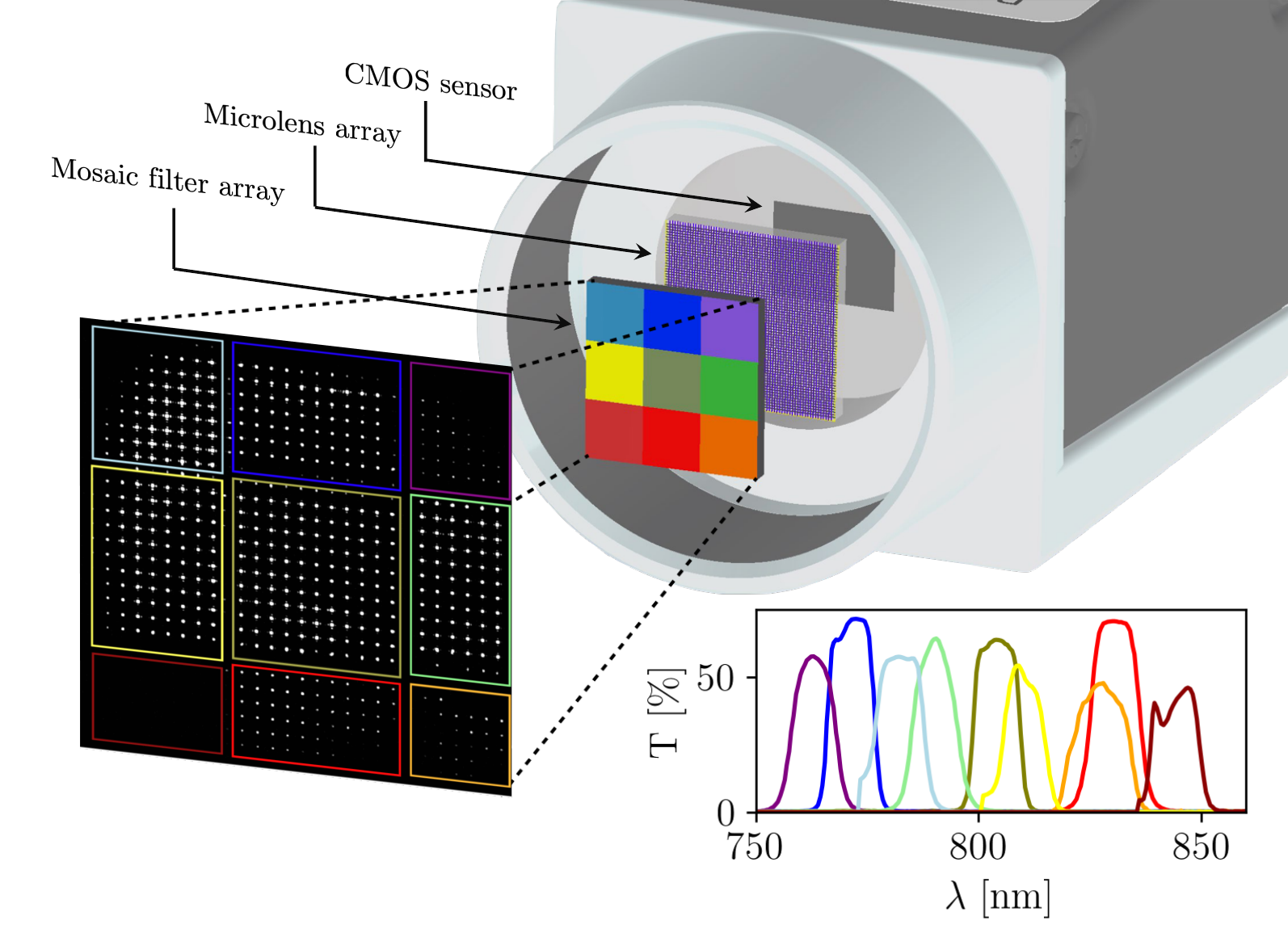}
    \caption{Single-shot FALCON, consisting of a mosaic bandpass filter array in front of a microlens array placed in effective focal length in front of the camera. The inset on the bottom right shows the corresponding spectral transmission of the mosaic filter array, measured with a spectrophotometer. Magnified is a raw image with filter positions marked in their respective colors. Note that the weaker signal behind the filters centered at around 765 and 845 nm is due to the spectrum of the ATLAS laser.}
    \label{fig:All_in_one}
\end{figure}

We now apply the Bayesian framework developed in the last section to the measurement of spatio-temporal couplings in a petawatt laser. We demonstrate single-shot sensitivity to spatio-temporal couplings of the ATLAS-3000 petawatt laser using a new, simple measurement device. 

\subsection{Single-shot measurements using mosaic filter array}

In \cite{weisse2023b} we introduced FALCON, an approach to reconstruct spatial temporal couplings based on measurements of the entire wavefront at frequencies selected with narrowband filters. While simple and fast compared to established schemes, the method requires on the order of $\sim 10-50$ shots, to (a) scan through different spectral filters and (b) to have enough statistics to suppress noise due to pointing jitter within the measurement set. 

To alleviate this problem, we introduce a single-shot variation of the technique that promises easy experimental implementation. The key idea is that the Zernike modes cover the full spatial domain of the measurement - as opposed to the pixel basis (which is often applied in the zonal reconstruction approach). This means that one can \textit{in principle} (given a noise-free environment) use a sub-aperture measurement to deduce the entire wavefront. By filtering each sub-aperture with different spectral filters, we can then measure spatio-temporal couplings. 

In this sub-aperture measurement approach, a mosaic filter array divides the wavefront into a series of sub-apertures \( A_i \) (where \( i \) indexes the sub-aperture), each with its own narrowband filter characterized by a central wavelength \( \omega_i \). The array consists of nine bandpass filters by Omega Optical, glued together into a $3\times3$ array \cite{haffa19}. The device is sketched in \cref{fig:All_in_one}. The nine bandpass filters cover the spectral range from $\SI{760}{\nano\metre}$ to $\SI{840}{\nano\metre}$, see bottom right of\cref{fig:All_in_one} for spectrophotometer measurements of their transmission curves. The arrangement allows us to measure the wavefront in different section of the beam for different colors. Our analysis shows that this configuration is well-suited to measure pulse front tilt and curvature, in particular.

To ensure the accuracy of this Single-Shot FALCON device it was validated against FALCON measurements in test scenarios with a broadband fs-oscillator. It should be noted that, before measuring with the Single-Shot FALCON, it is beneficial to measure the wavefront once to ensure that no high order spatial distortions are present that could be misinterpreted as lower order spatio-spectral effects. More specifically, if one uses a FALCON or similar device to measure the full spatio-spectral phase at the start, this can be used as a prior on the Single-Shot FALCON measurements instead of a flat phase.

Note that an alternative implementation of a single-shot device could be to use a beamsplitter e.g. a diffractive optical element (DOE) \cite{haffa19}, and to spectrally filter individual copies of the entire beam. However, the introduction of a DOE would increase complexity and cost of the setup. In contrast, the solution presented here is based on the addition of a single optical element, the mosaic filter, in front of an existing wavefront sensor, making it simple to implement and use.

\subsection{Experimental setup at the ATLAS-3000 Petawatt laser}
The ATLAS-3000 laser is a Titanium:Sapphire (Ti:Sa) laser system, capable of a maximum pulse energy of 90 Joules prior to compression. Post compression, the laser pulse has a duration of \SI{27}{\femto\second}, categorizing its peak power in the Petawatt regime. The beam diameter after the final expander measures \SI{27}{\centi\metre}.
For particle acceleration experiments, the laser pulse can be delivered with full energy or, for diagnostic purposes, significantly reduced energy by using a reflective attenuator placed behind the final amplifiers and before the laser beam expander and compressor. For the measurements presented in the following, the laser was operated at 15.0 Joules before attenuation and compression. 
Since particle acceleration experiments are highly sensitive to fluctuations in the laser intensity profile, a diagnostic that allows measurement of spatio-temporal couplings on a shot-to-shot basis is desirable. 

Measurements are performed on a de-magnified image of the laser's near field. For this purpose, we use a telescope consisting of a spherical mirror with $f_1=\SI{10}{\metre}$ focal length, otherwise used for electron acceleration experiments, combined with an $f_2=\SI{0.5}{\metre}$ achromatic doublet lens, followed by a 3:5 re-imaging and de-magnifying telescope. The laser enters the spherical mirror under a small angle, and the resulting aberrations are pre-corrected using a deformable mirror. The telescope images a plane of the near field approximately \SI{5}{\metre} before the spherical mirror. The intermediate focus of the telescope has a size of approximately \SI{30}{\micro\metre} diameter. To calibrate the phase measurement, we place a motorized pinhole with \SI{20}{\micro\metre} diameter at the focus position. The beam is fully imaged on the in-vacuum Shack-Hartmann sensor with a beam diameter of approximately \SI{8}{mm}, corresponding to a de-magnification of $1:34$.

\subsection{Bayesian modal reconstruction}\label{sec:modal}

As introduced in Ref.\cite{weisse2023b}, the spatio-spectral phase $\Phi(x,y,\omega)$ of a laser pulse can be expressed in terms of the Zernike-Taylor coefficients $a_{m,n}$, which are the equivalent of $x_k$ in the previous sections:
\begin{equation}
    \Phi(x,y,\omega) = \sum_{m,n,i}  a_{m,n}^i(\omega-\omega_0)^i \cdot Z_n^m(x,y),
    \label{eq:spatio-spectral-phase}
\end{equation}
where $a_{m,n}^i = \left({\partial^i a_{m,n}}/{\partial \omega
^i}\right)_{\omega = \omega_0}$, $\omega_0$ is the central frequency and $Z_n^m$ are the Zernike polynomials.

A frequency-resolving Shack-Hartmann wavefront sensor measures the gradient of the spatio-spectral phase, which can be related to coefficients describing spatio-temporal couplings through a transfer matrix $\mathbf{T}$:
\begin{equation}
    \vec{d_k} = \mathbf{T} \cdot \vec{a_k}+\vec{n},
    \label{eq:transfer_equation}
\end{equation}
where $\vec{d_k}$ is a vector containing the measured centroid positions and $\vec{a_k}$ is a vector of the retrieved coefficients for spatio-temporal couplings. $\vec{n}$ describes the noise of our measurement, which is assumed to be Gaussian. While traditional approaches solve this problem using least-squares formulation, the sub-aperture nature of our mosaic filter measurements introduces additional challenges. In a zonal reconstruction this would be completely impossible. In the modal approach, the transfer matrix becomes less well-conditioned compared to full-aperture measurements, as we are attempting to reconstruct global modes from a reduced set of local measurements. This poorer conditioning makes the system more sensitive to noise, potentially amplifying measurement uncertainties in the reconstruction. This is where our Bayesian framework proves especially valuable. By incorporating prior knowledge about the wavefront and properly accounting for measurement uncertainties, we can achieve robust reconstruction even with partial aperture measurements. The formalism outlined in the previous sections adapts naturally to the vectorial case of reconstructing multiple Zernike-Taylor coefficients at once through matrix equations, allowing us to properly weight the contribution of each sub-aperture measurement while maintaining the global coherence of the reconstruction. This approach transforms what might be an ill-conditioned inverse problem in a traditional least-squares framework into a well-posed probabilistic inference task.
The equivalent of \cref{eq:posterior_mean} becomes
\begin{equation}
    \vec{a}_{k|k} = \left(\mathbf{1}-\boldsymbol{\Gamma_k}\right) \vec{a}_{k|k-1} + \boldsymbol{\Gamma_k}\cdot \left( \mathbf{T^+} \vec{d}_k \right)
\end{equation}
where $\mathbf{T^+}$ is the pseudo-inverse of the transfer matrix. 
\cref{eq:posterior_variance} becomes 
\begin{equation}
    \boldsymbol{\Sigma^2_{k|k}} = \left( \mathbf{1}-\boldsymbol{\Gamma_k} \right) \boldsymbol{\Sigma^2_{k|k-1}} \left( \mathbf{1}-\boldsymbol{\Gamma_k} \right)^T + \boldsymbol{\Gamma_k} \boldsymbol{\Sigma^2_\text{meas}} \boldsymbol{\Gamma_k}^T
\end{equation}

and the update weight \cref{eq:kalman_gain} translates to
\begin{equation}
    \boldsymbol{\Gamma_k} = \boldsymbol{\Sigma^2_{k|k-1}}\cdot\left( \boldsymbol{\Sigma^2_{k|k-1}} +\boldsymbol{\Sigma^2_\text{meas}}\right)^{-1}.
\end{equation}
In this notation, bold capital $\boldsymbol{\Sigma}$ and $\boldsymbol{\Gamma}$ represent the matrix versions of the scalar variance $\sigma$ and update weight $\gamma$, respectively.

\begin{figure*}[]
    \centering
    \includegraphics[width=\linewidth]{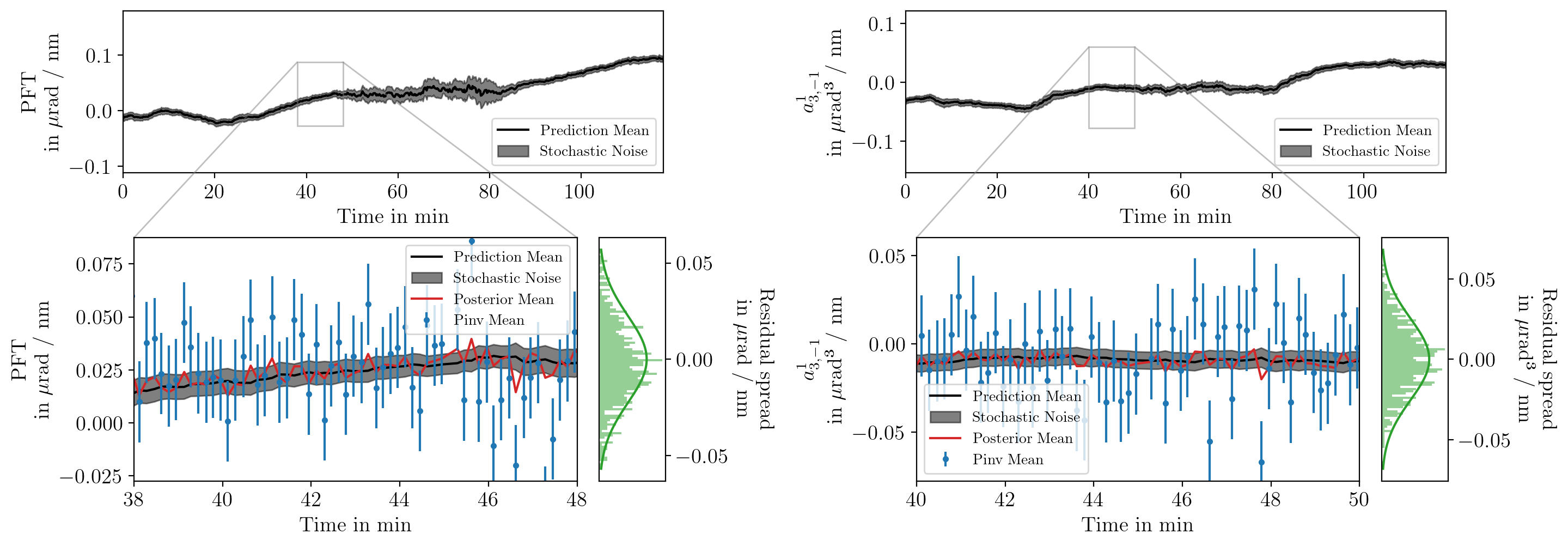}
    \caption{The evolution of the pulse front tilt (left) and linear-frequency dependent-coma (right) prediction values together with the intrinsic stochasticity. The zoomed sections add the posterior mean as well as the pseudo-inverse (Pinv) or least-squares mean. The histograms show the spread of the residuals as well as the equivalent Gaussian distribution obtained from our estimate \cref{eq:variance_residual}.}
    \label{fig:pft_plot}
\end{figure*}

\subsection{Calibration and estimating measurement noise}\label{sec:measurement-noise}

A key component to the above formalism is knowledge of the measurement uncertainty $\boldsymbol{\Sigma}^2_{meas}$. The matrix formalism allows us to calculate the variance on the coefficients given the uncertainty of the local phase gradients $\vec{n}$:
\begin{equation}
    \boldsymbol{\Sigma^2_\text{meas}} = \mathbf{T^+} \mathbf{n} \mathbf{T}^{+T}
\end{equation}
where $\mathbf{n}$ is the diagonal covariance matrix assigned to $\vec{n}$.
There are two types of uncertainty that have to be considered. For obvious reasons, the focii spot positions of a Shack-Hartmann can be determined only up to a certain precision due to their finite width on a sensor. We use a combination of a local 2D peak finder for coarse position determination followed by a 2D Gaussian fit for sub-pixel accuracy.
The second reason is that fitting a limited modal model to a zonal detector setup cannot include artifacts at high spatial frequencies. This could only be compensated by fitting high order Zernike polynomials, which showed to be numerically instable. Therefore, these high order fluctuations manifest as an uncorrelated noise term. This adds additional uncertainty to any fit in the pixel base domain.

For our measurements we used the Thorlabs MLA300-14AR microlens array and the IDS U3-3992SE camera with a pixel size of $\SI{2.74}{\micro\meter}$. From calibration measurements we estimate the accuracy in the focal spot position reconstruction to $0.2$ pixels. We assume this error for all well-illuminated pixels. While one might want to consider photon noise, its influence is on the focal spot estimate is difficult to estimate, so instead it was decided to remove individual microlens spots with insufficient illumination from the reconstruction set. We estimate the model fit uncertainty as an additional factor of $0.6$ pixels.
Note that this estimation of the measurement uncertainty is the first time this has been done for an STC diagnostic.

\subsection{Single Shot Measurements of STCs}\label{sec:results}

To demonstrate our diagnostic, we monitored the change of the STCs of the ATLAS-3000 Laser during continuous operation for about two hours. During this time, the laser system was not actively changed, but still subject to thermalization and other types of drifts.

We used a sampling frequency $\omega_s = \SI{0.1}{Hz}$ and a sliding window size $N=20$ to fit both the stochasticity and the local linear model. This covers about $\SI{3}{minutes}$. Parameter optimization yields, on average, $\bar\alpha=0.1596$ and $\bar\beta=0.0080$. These values are easier interpreted as "half-life" periods: the level component ($\alpha$) has a half-life of about $\SI{40}{s}$, meaning that the influence of any measurement on the estimated state decays to half its original value after this time. The trend component ($\beta$) has a much longer half-life of approximately $\SI{14.4}{min}$, indicating that -- on the timescale of the measurements -- trend changes slowly. The results for the prediction mean $\hat{x}$ and the intrinsic stochasticity $\sigma_\text{stoch}$ as uncertainty on that value are exemplary shown for the Pulse Front Tilt (PFT) and linear-frequency dependent-coma (linear coma) in \cref{fig:pft_plot}. 

It can be seen that there are dynamics on the longer timescales that go beyond the lasers intrinsic stochasticity. As discussed in \cref{sec:frequency-analysis}, if one would measure with a sampling rate at this timescale the fitted intrinsic stochasticity would be larger, as these dynamics would be incorporated into the stochasticity estimate.
To prove the legitimacy of these results, \cref{fig:pft_plot} also shows a ten minute section together with the results of the posterior mean as well as a result if only a least squares solution to the forward matrix -- calculated using a pseudo-inverse approach -- would have been used. It can be seen that the posterior mean follows the least squares results far more than the predicted mean but its fluctuations stay at the same order as the determined intrinsic stochasticity. This indicates that the predicted mean together with the intrinsic stochasticity accurately represents the intrinsic systems state with its uncertainty. The average improvement of the posterior uncertainty compared to the measurement uncertainty is about \SI{54}{\%} and \SI{60}{\%}, respectively. The histograms show that the residuals match the distribution (green curve) that one would expect from \cref{eq:variance_residual}. Additional plots for other modes can be found in the supplemental material.

\begin{figure}[]
    \centering
    \includegraphics[width=0.98\linewidth]{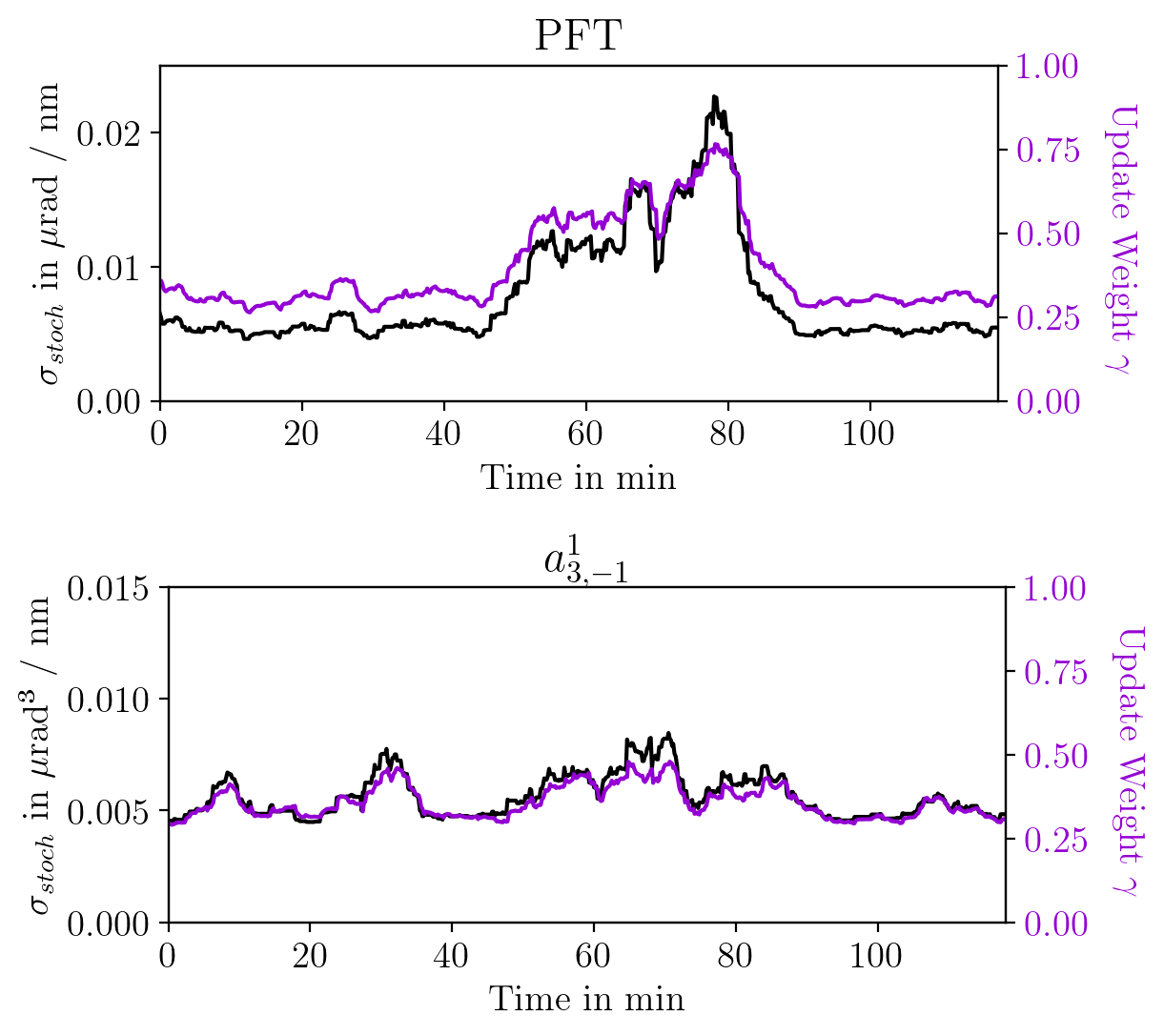}
    \caption{Update weight and intrinsic stochasticity for PFT (top) and linear coma (bottom).}
    \label{fig:Kalman_gain_plot}
\end{figure}

Additionally, one can look at the change of the update weight compared to the intrinsic stochasticity. The results are shown for PFT and linear coma in Fig. \ref{fig:Kalman_gain_plot}. From both it can be seen that the update weight is directly correlated to the intrinsic stochasticity. The update weights remain between bounds of 0.3-0.75 and 0.25-0.5 for PFT and linear coma, respectively. This implies that for the PFT with this measurement setup we are at the edge between a state-resolving and a distribution-centric regime. More precisely, according to \cref{eq:magnitude-response}, state components are attenuated by around 50\% in this setting. This implies that there is a limited, but non-zero advantage to having a 'single-shot' device with this measurement uncertainty looking. But a few- or multi-shot device can also be a reasonable choice. However, only this gives us an insight into the intrinsic stochasticity. For the linear coma we are entirely in the distribution-centric regime, whereas the reconstruction for the Tilt (shown in the supplemental) is entirely in the state-resolving regime. This demonstrates the more nuanced view provided by our framework.

For the PFT one can calculate the intensity loss in focus $\chi_\text{intensity}$ due to the intrinsic stochasticity on the PFT by the following analytic formula, adapted from \cite{pretzler2000angular}:
\begin{equation}
\chi_\text{intensity}= \left(1+\left( \frac{\sigma_\text{stoch}^\text{PFT}\sigma_\lambda D}{0.84\lambda_0}\right)\right)^{-2}
\end{equation}
with $\sigma_\lambda = \SI{30}{nm}$ being the bandwidth of the equivalent Gaussian spectrum, $D = \SI{27}{cm}$ the lasers diameter and $\lambda_0 = \SI{800}{nm}$. As $\sigma_\text{stoch}^\text{PFT}$ is between $0.005$ and $\SI{0.02}{\micro rad/nm}$, this corresponds to $0.3$ and $\SI{5.5}{\%}$ decrease in focus intensity. Hence, if the prediction mean of the PFT would be compensated for by adjusting the compressor grating angle, this would be the decrease compared to an ideal flat-top beam. As these fluctuations can not be compensated for due to their apparent stochastic nature, this provides a lower limit for peak intensity fluctuations of an otherwise perfect laser.

\begin{figure}[]
    \centering
    \includegraphics[width=\linewidth]{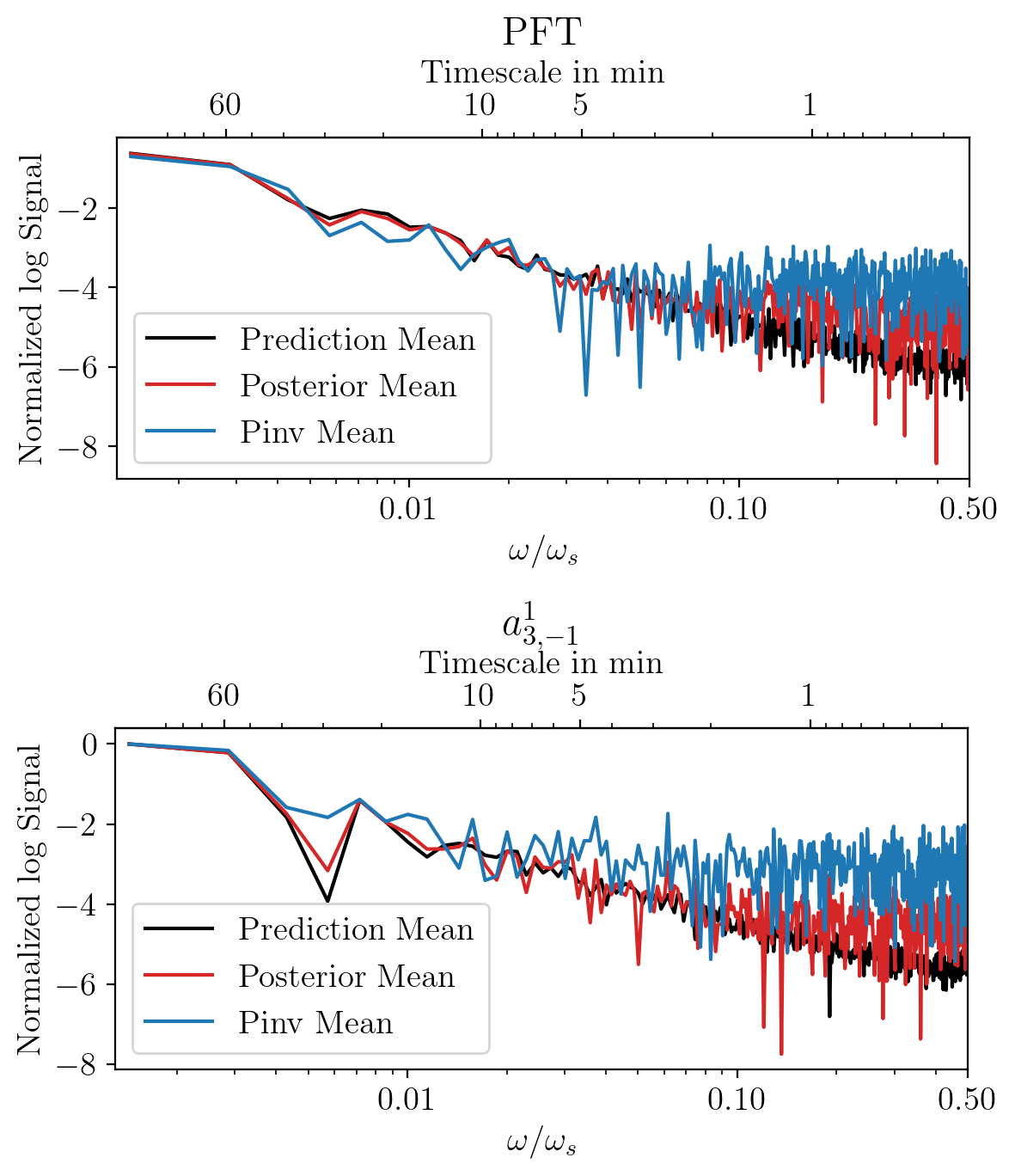}
    \caption{Logarithmic spectra for the prediction, posterior and pseudo-inverse mean of the PFT (top) and linear coma (bottom).}
    \label{fig:Spectral_response_plot}
\end{figure}

Lastly, Fig. \ref{fig:Spectral_response_plot} shows how the spectra for the predicted, posterior and pseudo-inverse mean of PFT and linear coma look like. It can be seen that for low sampling rates the three shown curves overlap whereas features at higher rates the pseudo-inverse mean gives the strongest signal and the predicted mean the lowest, as expected. The point, where the three curves separate gives, up to which timescales general trends are dominating and from when on statistical fluctuations dominate. For the PFT this is at approximately $\SI{8}{min}$ and for the linear coma between 40 and $\SI{20}{min}$. Notice also how the posterior's frequency spectrum is below the least square solution's for high frequency modulations, indicative of how the Bayesian update helps us to remove measurement noise while still resolving the intrinsic stochasticity of the system.


\section{Conclusions and Outlook}

In this work, we have introduced a novel perspective on laser characterization based on the idea of Bayesian inference. In this approach, the very concept of what constitutes a 'single-shot' measurement is challenged and instead, we argue that it is more suitable to distinguish between the capacity to resolve the average state of the system versus resolving stochastic shot-to-shot fluctuations. Along with this, we have developed a rigorous framework that handles Bayesian updates, as well as dynamic fitting of the system's underlying, predictable trends and unpredictable variability.

We have then applied the framework to the problem of monitoring spatio-temporal couplings in a petawatt laser system. By employing a modal reconstruction technique based on Zernike-Legendre polynomials and a mosaic filter array combined with a Shack-Hartmann wavefront sensor, we have demonstrated the retrieval of key STCs from single measurements. Crucially, the Bayesian framework allows us to obtain a posterior uncertainty that is well below the pure measurement uncertainty. These measurements constitute, to our knowledge, the first example for measurement uncertainty and shot-to-shot variability of spatio-temporal couplings. 

Future work will focus on extending the proposed approach to more complex spatio-temporal field structures. Our framework can both serve as input for predictive control algorithms for beam stabilization and as an estimate for the amount of fluctuations that cannot be corrected for. Crucially, while it might not be possible to correct for all changes \textit{a priori}, we can use the \textit{a posteriori} information from a state-resolving diagnostic to remove uncertainty from physical models. For instance, shot-to-shot information on pulse front tilt may serve as crucial input to refine probabilistic models of laser wakefield accelerators \cite{dopp2023data, PhysRevLett.133.085001}.

The framework itself can be developed further, for instance using other basis functions or using Gaussian processes to model both correlations in time and between variables. The availability of uncertainty estimates allows the development of new, adaptive diagnostic: The principles of Bayesian inference can be extended beyond data analysis to inform the design of experiments themselves. Bayesian experimental design aims to optimize measurement strategies to maximize information gain, which could be particularly valuable in the context of single-shot diagnostics where each measurement is precious. 

In conclusion, our work has introduced a new perspective on single-shot measurements of STCs, leveraging Bayesian inference and prior knowledge to redefine the concept. This paradigm shift opens up new possibilities for efficient and targeted optimization of high-power laser systems and their applications.

\section*{Acknowledgements}
This work was supported by the Independent Junior Research Group "Characterization and control of high-intensity laser pulses for particle acceleration", DFG Project No.~453619281. P.N. acknowledges support from UKRI-STFC grant ST/V001655/1.

\bibliography{references}

\begin{thebibliography}{31}
\expandafter\ifx\csname natexlab\endcsname\relax\def\natexlab#1{#1}\fi
\expandafter\ifx\csname bibnamefont\endcsname\relax
  \def\bibnamefont#1{#1}\fi
\expandafter\ifx\csname bibfnamefont\endcsname\relax
  \def\bibfnamefont#1{#1}\fi
\expandafter\ifx\csname citenamefont\endcsname\relax
  \def\citenamefont#1{#1}\fi
\expandafter\ifx\csname url\endcsname\relax
  \def\url#1{\texttt{#1}}\fi
\expandafter\ifx\csname urlprefix\endcsname\relax\def\urlprefix{URL }\fi
\providecommand{\bibinfo}[2]{#2}
\providecommand{\eprint}[2][]{\url{#2}}

\bibitem[{\citenamefont{Danson et~al.}(2015)\citenamefont{Danson, Hillier,
  Hopps, and Neely}}]{Danson.2015}
\bibinfo{author}{\bibfnamefont{C.}~\bibnamefont{Danson}},
  \bibinfo{author}{\bibfnamefont{D.}~\bibnamefont{Hillier}},
  \bibinfo{author}{\bibfnamefont{N.}~\bibnamefont{Hopps}}, \bibnamefont{and}
  \bibinfo{author}{\bibfnamefont{D.}~\bibnamefont{Neely}},
  \bibinfo{journal}{High Power Laser Science and Engineering}
  \textbf{\bibinfo{volume}{3}}, \bibinfo{pages}{e3} (\bibinfo{year}{2015}).

\bibitem[{\citenamefont{Yoon et~al.}(2021)\citenamefont{Yoon, Kim, Choi, Sung,
  Lee, Lee, and Nam}}]{yoon2021realization}
\bibinfo{author}{\bibfnamefont{J.~W.} \bibnamefont{Yoon}},
  \bibinfo{author}{\bibfnamefont{Y.~G.} \bibnamefont{Kim}},
  \bibinfo{author}{\bibfnamefont{I.~W.} \bibnamefont{Choi}},
  \bibinfo{author}{\bibfnamefont{J.~H.} \bibnamefont{Sung}},
  \bibinfo{author}{\bibfnamefont{H.~W.} \bibnamefont{Lee}},
  \bibinfo{author}{\bibfnamefont{S.~K.} \bibnamefont{Lee}}, \bibnamefont{and}
  \bibinfo{author}{\bibfnamefont{C.~H.} \bibnamefont{Nam}},
  \bibinfo{journal}{Optica} \textbf{\bibinfo{volume}{8}}, \bibinfo{pages}{630}
  (\bibinfo{year}{2021}).

\bibitem[{\citenamefont{Bowlan et~al.}(2007)\citenamefont{Bowlan, Gabolde, and
  Trebino}}]{bowlan07}
\bibinfo{author}{\bibfnamefont{P.}~\bibnamefont{Bowlan}},
  \bibinfo{author}{\bibfnamefont{P.}~\bibnamefont{Gabolde}}, \bibnamefont{and}
  \bibinfo{author}{\bibfnamefont{R.}~\bibnamefont{Trebino}},
  \bibinfo{journal}{Optics Express} \textbf{\bibinfo{volume}{15}},
  \bibinfo{pages}{10219} (\bibinfo{year}{2007}).

\bibitem[{\citenamefont{Bowlan et~al.}(2008)\citenamefont{Bowlan, Gabolde,
  Coughlan, Trebino, and Levis}}]{bowlan08}
\bibinfo{author}{\bibfnamefont{P.}~\bibnamefont{Bowlan}},
  \bibinfo{author}{\bibfnamefont{P.}~\bibnamefont{Gabolde}},
  \bibinfo{author}{\bibfnamefont{M.~A.} \bibnamefont{Coughlan}},
  \bibinfo{author}{\bibfnamefont{R.}~\bibnamefont{Trebino}}, \bibnamefont{and}
  \bibinfo{author}{\bibfnamefont{R.~J.} \bibnamefont{Levis}},
  \bibinfo{journal}{Journal of the Optical Society of America B}
  \textbf{\bibinfo{volume}{25}}, \bibinfo{pages}{A81} (\bibinfo{year}{2008}).

\bibitem[{\citenamefont{Gallet et~al.}(2014)\citenamefont{Gallet, Kahaly,
  Gobert, and Qu{\'e}r{\'e}}}]{gallet14-2}
\bibinfo{author}{\bibfnamefont{V.}~\bibnamefont{Gallet}},
  \bibinfo{author}{\bibfnamefont{S.}~\bibnamefont{Kahaly}},
  \bibinfo{author}{\bibfnamefont{O.}~\bibnamefont{Gobert}}, \bibnamefont{and}
  \bibinfo{author}{\bibfnamefont{F.}~\bibnamefont{Qu{\'e}r{\'e}}},
  \bibinfo{journal}{Optics Letters} \textbf{\bibinfo{volume}{39}},
  \bibinfo{pages}{4687} (\bibinfo{year}{2014}).

\bibitem[{\citenamefont{Alonso et~al.}(2010)\citenamefont{Alonso, Sola, Varela,
  Hern{\'a}ndez-Toro, M{\'e}ndez, Rom{\'a}n, Za{\"i}r, and Roso}}]{alonso10}
\bibinfo{author}{\bibfnamefont{B.}~\bibnamefont{Alonso}},
  \bibinfo{author}{\bibfnamefont{{\'I}.~J.} \bibnamefont{Sola}},
  \bibinfo{author}{\bibfnamefont{{\'O}.}~\bibnamefont{Varela}},
  \bibinfo{author}{\bibfnamefont{J.}~\bibnamefont{Hern{\'a}ndez-Toro}},
  \bibinfo{author}{\bibfnamefont{C.}~\bibnamefont{M{\'e}ndez}},
  \bibinfo{author}{\bibfnamefont{J.~S.} \bibnamefont{Rom{\'a}n}},
  \bibinfo{author}{\bibfnamefont{A.}~\bibnamefont{Za{\"i}r}}, \bibnamefont{and}
  \bibinfo{author}{\bibfnamefont{L.}~\bibnamefont{Roso}},
  \bibinfo{journal}{Journal of the Optical Society of America B}
  \textbf{\bibinfo{volume}{27}}, \bibinfo{pages}{933} (\bibinfo{year}{2010}).

\bibitem[{\citenamefont{Alonso et~al.}(2012{\natexlab{a}})\citenamefont{Alonso,
  Miranda, Sola, and Crespo}}]{alonso12-1}
\bibinfo{author}{\bibfnamefont{B.}~\bibnamefont{Alonso}},
  \bibinfo{author}{\bibfnamefont{M.}~\bibnamefont{Miranda}},
  \bibinfo{author}{\bibfnamefont{{\'I}.~J.} \bibnamefont{Sola}},
  \bibnamefont{and} \bibinfo{author}{\bibfnamefont{H.}~\bibnamefont{Crespo}},
  \bibinfo{journal}{Optics Express} \textbf{\bibinfo{volume}{20}},
  \bibinfo{pages}{17880} (\bibinfo{year}{2012}{\natexlab{a}}).

\bibitem[{\citenamefont{Alonso et~al.}(2012{\natexlab{b}})\citenamefont{Alonso,
  Borrego-Varillas, Mendoza-Yero, Sola, Rom{\'a}n, M{\'i}nguez-Vega, and
  Roso}}]{alonso12-2}
\bibinfo{author}{\bibfnamefont{B.}~\bibnamefont{Alonso}},
  \bibinfo{author}{\bibfnamefont{R.}~\bibnamefont{Borrego-Varillas}},
  \bibinfo{author}{\bibfnamefont{O.}~\bibnamefont{Mendoza-Yero}},
  \bibinfo{author}{\bibfnamefont{{\'I}.~J.} \bibnamefont{Sola}},
  \bibinfo{author}{\bibfnamefont{J.~S.} \bibnamefont{Rom{\'a}n}},
  \bibinfo{author}{\bibfnamefont{G.}~\bibnamefont{M{\'i}nguez-Vega}},
  \bibnamefont{and} \bibinfo{author}{\bibfnamefont{L.}~\bibnamefont{Roso}},
  \bibinfo{journal}{Journal of the Optical Society of America B}
  \textbf{\bibinfo{volume}{29}}, \bibinfo{pages}{1993}
  (\bibinfo{year}{2012}{\natexlab{b}}).

\bibitem[{\citenamefont{Alonso et~al.}(2013)\citenamefont{Alonso, Miranda,
  Silva, Pervak, Rauschenberger, Rom{\'a}n, Sola, and Crespo}}]{alonso13}
\bibinfo{author}{\bibfnamefont{B.}~\bibnamefont{Alonso}},
  \bibinfo{author}{\bibfnamefont{M.}~\bibnamefont{Miranda}},
  \bibinfo{author}{\bibfnamefont{F.}~\bibnamefont{Silva}},
  \bibinfo{author}{\bibfnamefont{V.}~\bibnamefont{Pervak}},
  \bibinfo{author}{\bibfnamefont{J.}~\bibnamefont{Rauschenberger}},
  \bibinfo{author}{\bibfnamefont{J.~S.} \bibnamefont{Rom{\'a}n}},
  \bibinfo{author}{\bibfnamefont{{\'I}.~J.} \bibnamefont{Sola}},
  \bibnamefont{and} \bibinfo{author}{\bibfnamefont{H.}~\bibnamefont{Crespo}},
  \bibinfo{journal}{Applied Physics B} \textbf{\bibinfo{volume}{112}},
  \bibinfo{pages}{105} (\bibinfo{year}{2013}).

\bibitem[{\citenamefont{Miranda et~al.}(2014)\citenamefont{Miranda, Kotur,
  Rudawski, Guo, Harth, L’Huillier, and Arnold}}]{miranda14}
\bibinfo{author}{\bibfnamefont{M.}~\bibnamefont{Miranda}},
  \bibinfo{author}{\bibfnamefont{M.}~\bibnamefont{Kotur}},
  \bibinfo{author}{\bibfnamefont{P.}~\bibnamefont{Rudawski}},
  \bibinfo{author}{\bibfnamefont{C.}~\bibnamefont{Guo}},
  \bibinfo{author}{\bibfnamefont{A.}~\bibnamefont{Harth}},
  \bibinfo{author}{\bibfnamefont{A.}~\bibnamefont{L’Huillier}},
  \bibnamefont{and} \bibinfo{author}{\bibfnamefont{C.~L.}
  \bibnamefont{Arnold}}, \bibinfo{journal}{Optics Letters}
  \textbf{\bibinfo{volume}{39}}, \bibinfo{pages}{5142} (\bibinfo{year}{2014}).

\bibitem[{\citenamefont{Pariente et~al.}(2016)\citenamefont{Pariente, Gallet,
  Borot, Gobert, and Qu{\'e}r{\'e}}}]{pariente16}
\bibinfo{author}{\bibfnamefont{G.}~\bibnamefont{Pariente}},
  \bibinfo{author}{\bibfnamefont{V.}~\bibnamefont{Gallet}},
  \bibinfo{author}{\bibfnamefont{A.}~\bibnamefont{Borot}},
  \bibinfo{author}{\bibfnamefont{O.}~\bibnamefont{Gobert}}, \bibnamefont{and}
  \bibinfo{author}{\bibfnamefont{F.}~\bibnamefont{Qu{\'e}r{\'e}}},
  \bibinfo{journal}{Nature Photonics} \textbf{\bibinfo{volume}{10}},
  \bibinfo{pages}{547} (\bibinfo{year}{2016}).

\bibitem[{\citenamefont{Borot and Qu{\'e}r{\'e}}(2018)}]{borot2018spatio}
\bibinfo{author}{\bibfnamefont{A.}~\bibnamefont{Borot}} \bibnamefont{and}
  \bibinfo{author}{\bibfnamefont{F.}~\bibnamefont{Qu{\'e}r{\'e}}},
  \bibinfo{journal}{Optics express} \textbf{\bibinfo{volume}{26}},
  \bibinfo{pages}{26444} (\bibinfo{year}{2018}).

\bibitem[{\citenamefont{Weisse et~al.}(2023)\citenamefont{Weisse, Esslinger,
  Howard, Foerster, Haberstroh, Doyle, Norreys, Schreiber, Karsch, and
  D{\"o}pp}}]{weisse2023b}
\bibinfo{author}{\bibfnamefont{N.}~\bibnamefont{Weisse}},
  \bibinfo{author}{\bibfnamefont{J.}~\bibnamefont{Esslinger}},
  \bibinfo{author}{\bibfnamefont{S.}~\bibnamefont{Howard}},
  \bibinfo{author}{\bibfnamefont{F.~M.} \bibnamefont{Foerster}},
  \bibinfo{author}{\bibfnamefont{F.}~\bibnamefont{Haberstroh}},
  \bibinfo{author}{\bibfnamefont{L.}~\bibnamefont{Doyle}},
  \bibinfo{author}{\bibfnamefont{P.}~\bibnamefont{Norreys}},
  \bibinfo{author}{\bibfnamefont{J.}~\bibnamefont{Schreiber}},
  \bibinfo{author}{\bibfnamefont{S.}~\bibnamefont{Karsch}}, \bibnamefont{and}
  \bibinfo{author}{\bibfnamefont{A.}~\bibnamefont{D{\"o}pp}},
  \bibinfo{journal}{Optics Express} \textbf{\bibinfo{volume}{31}},
  \bibinfo{pages}{19733} (\bibinfo{year}{2023}), ISSN
  \bibinfo{issn}{1094-4087}.

\bibitem[{\citenamefont{Smartsev et~al.}(2024)\citenamefont{Smartsev, Liberman,
  Andriyash, Cavagna, Flacco, Giaccaglia, Kaur, Monzac, Tata, Vernier
  et~al.}}]{smartsev24}
\bibinfo{author}{\bibfnamefont{S.}~\bibnamefont{Smartsev}},
  \bibinfo{author}{\bibfnamefont{A.}~\bibnamefont{Liberman}},
  \bibinfo{author}{\bibfnamefont{I.}~\bibnamefont{Andriyash}},
  \bibinfo{author}{\bibfnamefont{A.}~\bibnamefont{Cavagna}},
  \bibinfo{author}{\bibfnamefont{A.}~\bibnamefont{Flacco}},
  \bibinfo{author}{\bibfnamefont{C.}~\bibnamefont{Giaccaglia}},
  \bibinfo{author}{\bibfnamefont{J.}~\bibnamefont{Kaur}},
  \bibinfo{author}{\bibfnamefont{J.}~\bibnamefont{Monzac}},
  \bibinfo{author}{\bibfnamefont{S.}~\bibnamefont{Tata}},
  \bibinfo{author}{\bibfnamefont{A.}~\bibnamefont{Vernier}},
  \bibnamefont{et~al.}, \bibinfo{journal}{Optics Letters}
  \textbf{\bibinfo{volume}{49}}, \bibinfo{pages}{1900} (\bibinfo{year}{2024}).

\bibitem[{\citenamefont{Gabolde and Trebino}(2006)}]{gabolde06}
\bibinfo{author}{\bibfnamefont{P.}~\bibnamefont{Gabolde}} \bibnamefont{and}
  \bibinfo{author}{\bibfnamefont{R.}~\bibnamefont{Trebino}},
  \bibinfo{journal}{Optics Express} \textbf{\bibinfo{volume}{14}},
  \bibinfo{pages}{11460} (\bibinfo{year}{2006}).

\bibitem[{\citenamefont{Dorrer and Bahk}(2018)}]{dorrer18}
\bibinfo{author}{\bibfnamefont{C.}~\bibnamefont{Dorrer}} \bibnamefont{and}
  \bibinfo{author}{\bibfnamefont{S.-W.} \bibnamefont{Bahk}},
  \bibinfo{journal}{Optics Express} \textbf{\bibinfo{volume}{26}},
  \bibinfo{pages}{33387} (\bibinfo{year}{2018}).

\bibitem[{\citenamefont{Kim et~al.}(2021)\citenamefont{Kim, Kim, Yoon, Sung,
  Lee, and Nam}}]{kim21}
\bibinfo{author}{\bibfnamefont{Y.~G.} \bibnamefont{Kim}},
  \bibinfo{author}{\bibfnamefont{J.~I.} \bibnamefont{Kim}},
  \bibinfo{author}{\bibfnamefont{J.~W.} \bibnamefont{Yoon}},
  \bibinfo{author}{\bibfnamefont{J.~H.} \bibnamefont{Sung}},
  \bibinfo{author}{\bibfnamefont{S.~K.} \bibnamefont{Lee}}, \bibnamefont{and}
  \bibinfo{author}{\bibfnamefont{C.~H.} \bibnamefont{Nam}},
  \bibinfo{journal}{Optics Express} \textbf{\bibinfo{volume}{29}},
  \bibinfo{pages}{19506} (\bibinfo{year}{2021}).

\bibitem[{\citenamefont{Tang et~al.}(2022)\citenamefont{Tang, Men, Liu, Hu, Su,
  Zuo, Li, Liang, Downer, and Li}}]{tang2022}
\bibinfo{author}{\bibfnamefont{H.}~\bibnamefont{Tang}},
  \bibinfo{author}{\bibfnamefont{T.}~\bibnamefont{Men}},
  \bibinfo{author}{\bibfnamefont{X.}~\bibnamefont{Liu}},
  \bibinfo{author}{\bibfnamefont{Y.}~\bibnamefont{Hu}},
  \bibinfo{author}{\bibfnamefont{J.}~\bibnamefont{Su}},
  \bibinfo{author}{\bibfnamefont{Y.}~\bibnamefont{Zuo}},
  \bibinfo{author}{\bibfnamefont{P.}~\bibnamefont{Li}},
  \bibinfo{author}{\bibfnamefont{J.}~\bibnamefont{Liang}},
  \bibinfo{author}{\bibfnamefont{M.~C.} \bibnamefont{Downer}},
  \bibnamefont{and} \bibinfo{author}{\bibfnamefont{Z.}~\bibnamefont{Li}},
  \bibinfo{journal}{Light: Science \& Applications}
  \textbf{\bibinfo{volume}{11}}, \bibinfo{pages}{244} (\bibinfo{year}{2022}),
  ISSN \bibinfo{issn}{2047-7538}.

\bibitem[{\citenamefont{Howard et~al.}(2023)\citenamefont{Howard, Esslinger,
  Wang, Norreys, and Döpp}}]{howard23}
\bibinfo{author}{\bibfnamefont{S.}~\bibnamefont{Howard}},
  \bibinfo{author}{\bibfnamefont{J.}~\bibnamefont{Esslinger}},
  \bibinfo{author}{\bibfnamefont{R.~H.~W.} \bibnamefont{Wang}},
  \bibinfo{author}{\bibfnamefont{P.}~\bibnamefont{Norreys}}, \bibnamefont{and}
  \bibinfo{author}{\bibfnamefont{A.}~\bibnamefont{Döpp}},
  \bibinfo{journal}{High Power Laser Science and Engineering}
  \textbf{\bibinfo{volume}{11}}, \bibinfo{pages}{e32} (\bibinfo{year}{2023}).

\bibitem[{\citenamefont{Alonso et~al.}(2024)\citenamefont{Alonso, D{\"o}pp, and
  Jolly}}]{alonso2024space}
\bibinfo{author}{\bibfnamefont{B.}~\bibnamefont{Alonso}},
  \bibinfo{author}{\bibfnamefont{A.}~\bibnamefont{D{\"o}pp}}, \bibnamefont{and}
  \bibinfo{author}{\bibfnamefont{S.~W.} \bibnamefont{Jolly}},
  \bibinfo{journal}{APL Photonics} \textbf{\bibinfo{volume}{9}}
  (\bibinfo{year}{2024}).

\bibitem[{\citenamefont{D{\"o}pp et~al.}(2023)\citenamefont{D{\"o}pp, Eberle,
  Howard, Irshad, Lin, and Streeter}}]{dopp2023data}
\bibinfo{author}{\bibfnamefont{A.}~\bibnamefont{D{\"o}pp}},
  \bibinfo{author}{\bibfnamefont{C.}~\bibnamefont{Eberle}},
  \bibinfo{author}{\bibfnamefont{S.}~\bibnamefont{Howard}},
  \bibinfo{author}{\bibfnamefont{F.}~\bibnamefont{Irshad}},
  \bibinfo{author}{\bibfnamefont{J.}~\bibnamefont{Lin}}, \bibnamefont{and}
  \bibinfo{author}{\bibfnamefont{M.}~\bibnamefont{Streeter}},
  \bibinfo{journal}{High Power Laser Science and Engineering}
  \textbf{\bibinfo{volume}{11}}, \bibinfo{pages}{e55} (\bibinfo{year}{2023}).

\bibitem[{\citenamefont{Debus et~al.}(2019)\citenamefont{Debus, Pausch, Huebl,
  Steiniger, Widera, Cowan, Schramm, and Bussmann}}]{debus19}
\bibinfo{author}{\bibfnamefont{A.}~\bibnamefont{Debus}},
  \bibinfo{author}{\bibfnamefont{R.}~\bibnamefont{Pausch}},
  \bibinfo{author}{\bibfnamefont{A.}~\bibnamefont{Huebl}},
  \bibinfo{author}{\bibfnamefont{K.}~\bibnamefont{Steiniger}},
  \bibinfo{author}{\bibfnamefont{R.}~\bibnamefont{Widera}},
  \bibinfo{author}{\bibfnamefont{T.~E.} \bibnamefont{Cowan}},
  \bibinfo{author}{\bibfnamefont{U.}~\bibnamefont{Schramm}}, \bibnamefont{and}
  \bibinfo{author}{\bibfnamefont{M.}~\bibnamefont{Bussmann}},
  \bibinfo{journal}{Physical Review X} \textbf{\bibinfo{volume}{9}},
  \bibinfo{pages}{031044} (\bibinfo{year}{2019}).

\bibitem[{\citenamefont{Palastro et~al.}(2020)\citenamefont{Palastro, Shaw,
  Franke, Ramsey, Simpson, and Froula}}]{palastro20}
\bibinfo{author}{\bibfnamefont{J.~P.} \bibnamefont{Palastro}},
  \bibinfo{author}{\bibfnamefont{J.~L.} \bibnamefont{Shaw}},
  \bibinfo{author}{\bibfnamefont{P.}~\bibnamefont{Franke}},
  \bibinfo{author}{\bibfnamefont{D.}~\bibnamefont{Ramsey}},
  \bibinfo{author}{\bibfnamefont{T.~T.} \bibnamefont{Simpson}},
  \bibnamefont{and} \bibinfo{author}{\bibfnamefont{D.~H.}
  \bibnamefont{Froula}}, \bibinfo{journal}{Physical Review Letters}
  \textbf{\bibinfo{volume}{124}}, \bibinfo{pages}{134802}
  (\bibinfo{year}{2020}).

\bibitem[{\citenamefont{Caizergues et~al.}(2020)\citenamefont{Caizergues,
  Smartsev, Malka, and Thaury}}]{caizergues20}
\bibinfo{author}{\bibfnamefont{C.}~\bibnamefont{Caizergues}},
  \bibinfo{author}{\bibfnamefont{S.}~\bibnamefont{Smartsev}},
  \bibinfo{author}{\bibfnamefont{V.}~\bibnamefont{Malka}}, \bibnamefont{and}
  \bibinfo{author}{\bibfnamefont{C.}~\bibnamefont{Thaury}},
  \bibinfo{journal}{Nature Photonics} \textbf{\bibinfo{volume}{14}},
  \bibinfo{pages}{475} (\bibinfo{year}{2020}).

\bibitem[{\citenamefont{Vincenti and Qu{\'e}r{\'e}}(2012)}]{vincenti12}
\bibinfo{author}{\bibfnamefont{H.}~\bibnamefont{Vincenti}} \bibnamefont{and}
  \bibinfo{author}{\bibfnamefont{F.}~\bibnamefont{Qu{\'e}r{\'e}}},
  \bibinfo{journal}{Physical Review Letters} \textbf{\bibinfo{volume}{108}},
  \bibinfo{pages}{113904} (\bibinfo{year}{2012}).

\bibitem[{\citenamefont{Qu{\'e}r{\'e} et~al.}(2014)\citenamefont{Qu{\'e}r{\'e},
  Vincenti, Borot, Monchoc{\'e}, Hammond, Kim, Wheeler, Zhang, Ruchon, Auguste
  et~al.}}]{quere14}
\bibinfo{author}{\bibfnamefont{F.}~\bibnamefont{Qu{\'e}r{\'e}}},
  \bibinfo{author}{\bibfnamefont{H.}~\bibnamefont{Vincenti}},
  \bibinfo{author}{\bibfnamefont{A.}~\bibnamefont{Borot}},
  \bibinfo{author}{\bibfnamefont{S.}~\bibnamefont{Monchoc{\'e}}},
  \bibinfo{author}{\bibfnamefont{T.~J.} \bibnamefont{Hammond}},
  \bibinfo{author}{\bibfnamefont{K.~T.} \bibnamefont{Kim}},
  \bibinfo{author}{\bibfnamefont{J.~A.} \bibnamefont{Wheeler}},
  \bibinfo{author}{\bibfnamefont{C.}~\bibnamefont{Zhang}},
  \bibinfo{author}{\bibfnamefont{T.}~\bibnamefont{Ruchon}},
  \bibinfo{author}{\bibfnamefont{T.}~\bibnamefont{Auguste}},
  \bibnamefont{et~al.}, \bibinfo{journal}{Journal of Physics B: Atomic,
  Molecular and Optical Physics} \textbf{\bibinfo{volume}{47}},
  \bibinfo{pages}{124004} (\bibinfo{year}{2014}).

\bibitem[{\citenamefont{Mohamed and Schwarz}(1999)}]{mohamed1999adaptive}
\bibinfo{author}{\bibfnamefont{A.}~\bibnamefont{Mohamed}} \bibnamefont{and}
  \bibinfo{author}{\bibfnamefont{K.}~\bibnamefont{Schwarz}},
  \bibinfo{journal}{Journal of geodesy} \textbf{\bibinfo{volume}{73}},
  \bibinfo{pages}{193} (\bibinfo{year}{1999}).

\bibitem[{\citenamefont{Holt}(2004)}]{holt2004forecasting}
\bibinfo{author}{\bibfnamefont{C.~C.} \bibnamefont{Holt}},
  \bibinfo{journal}{International journal of forecasting}
  \textbf{\bibinfo{volume}{20}}, \bibinfo{pages}{5} (\bibinfo{year}{2004}).

\bibitem[{\citenamefont{Haffa et~al.}(2019)\citenamefont{Haffa, Bin, Speicher,
  Allinger, Hartmann, Kreuzer, Ridente, Ostermayr, and Schreiber}}]{haffa19}
\bibinfo{author}{\bibfnamefont{D.}~\bibnamefont{Haffa}},
  \bibinfo{author}{\bibfnamefont{J.}~\bibnamefont{Bin}},
  \bibinfo{author}{\bibfnamefont{M.}~\bibnamefont{Speicher}},
  \bibinfo{author}{\bibfnamefont{K.}~\bibnamefont{Allinger}},
  \bibinfo{author}{\bibfnamefont{J.}~\bibnamefont{Hartmann}},
  \bibinfo{author}{\bibfnamefont{C.}~\bibnamefont{Kreuzer}},
  \bibinfo{author}{\bibfnamefont{E.}~\bibnamefont{Ridente}},
  \bibinfo{author}{\bibfnamefont{T.~M.} \bibnamefont{Ostermayr}},
  \bibnamefont{and}
  \bibinfo{author}{\bibfnamefont{J.}~\bibnamefont{Schreiber}},
  \bibinfo{journal}{Scientific Reports} \textbf{\bibinfo{volume}{9}},
  \bibinfo{pages}{7697} (\bibinfo{year}{2019}).

\bibitem[{\citenamefont{Pretzler et~al.}(2000)\citenamefont{Pretzler, Kasper,
  and Witte}}]{pretzler2000angular}
\bibinfo{author}{\bibfnamefont{G.}~\bibnamefont{Pretzler}},
  \bibinfo{author}{\bibfnamefont{A.}~\bibnamefont{Kasper}}, \bibnamefont{and}
  \bibinfo{author}{\bibfnamefont{K.}~\bibnamefont{Witte}},
  \bibinfo{journal}{Applied Physics B} \textbf{\bibinfo{volume}{70}},
  \bibinfo{pages}{1} (\bibinfo{year}{2000}).

\bibitem[{\citenamefont{Irshad et~al.}(2024)\citenamefont{Irshad, Eberle,
  Foerster, Grafenstein, Haberstroh, Travac, Weisse, Karsch, and
  D\"opp}}]{PhysRevLett.133.085001}
\bibinfo{author}{\bibfnamefont{F.}~\bibnamefont{Irshad}},
  \bibinfo{author}{\bibfnamefont{C.}~\bibnamefont{Eberle}},
  \bibinfo{author}{\bibfnamefont{F.~M.} \bibnamefont{Foerster}},
  \bibinfo{author}{\bibfnamefont{K.~v.} \bibnamefont{Grafenstein}},
  \bibinfo{author}{\bibfnamefont{F.}~\bibnamefont{Haberstroh}},
  \bibinfo{author}{\bibfnamefont{E.}~\bibnamefont{Travac}},
  \bibinfo{author}{\bibfnamefont{N.}~\bibnamefont{Weisse}},
  \bibinfo{author}{\bibfnamefont{S.}~\bibnamefont{Karsch}}, \bibnamefont{and}
  \bibinfo{author}{\bibfnamefont{A.}~\bibnamefont{D\"opp}},
  \bibinfo{journal}{Phys. Rev. Lett.} \textbf{\bibinfo{volume}{133}},
  \bibinfo{pages}{085001} (\bibinfo{year}{2024}),
  \urlprefix\url{https://link.aps.org/doi/10.1103/PhysRevLett.133.085001}.

\end{thebibliography}

\onecolumngrid
\newpage
\section*{Supplemental Material: Additional Modal Plots}

\begin{figure*}[h]
    \centering
    \includegraphics[width=\linewidth]{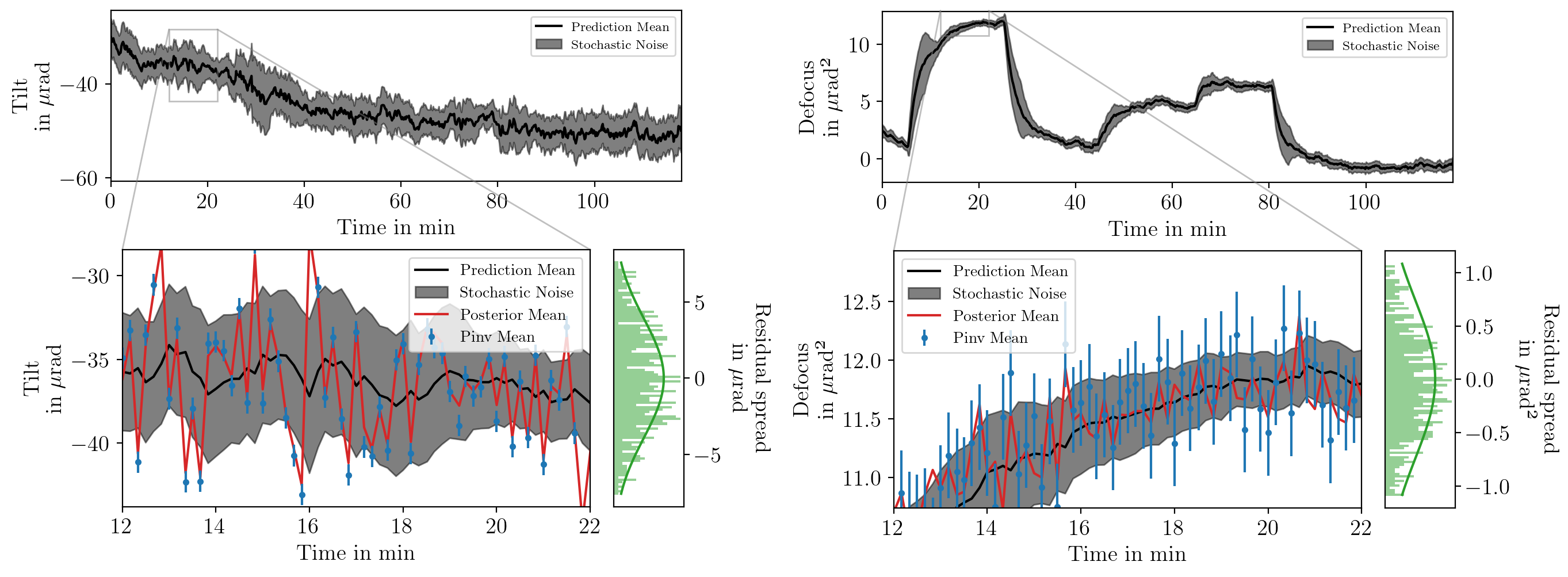}
    \caption{The evolution of the tilt (left) and defocus (right) prediction values together with the intrinsic stochasticity. The zoomed sections add the posterior mean as well as the pseudo-inverse (Pinv) or least-squares mean. The histograms show the spread of the residuals as well as the equivalent Gaussian distribution obtained from our estimate in Eq. 11. Respective uncertainty improvements are \SI{48}{\%} and \SI{59}{\%}. For the Tilt one can see that our framework functions equally well if the measurement uncertainty is lower than the stochastic noise.}
    \label{fig:linear_order_plot}
    
\end{figure*}

\begin{figure*}[h]
    \centering
    \includegraphics[width=\linewidth]{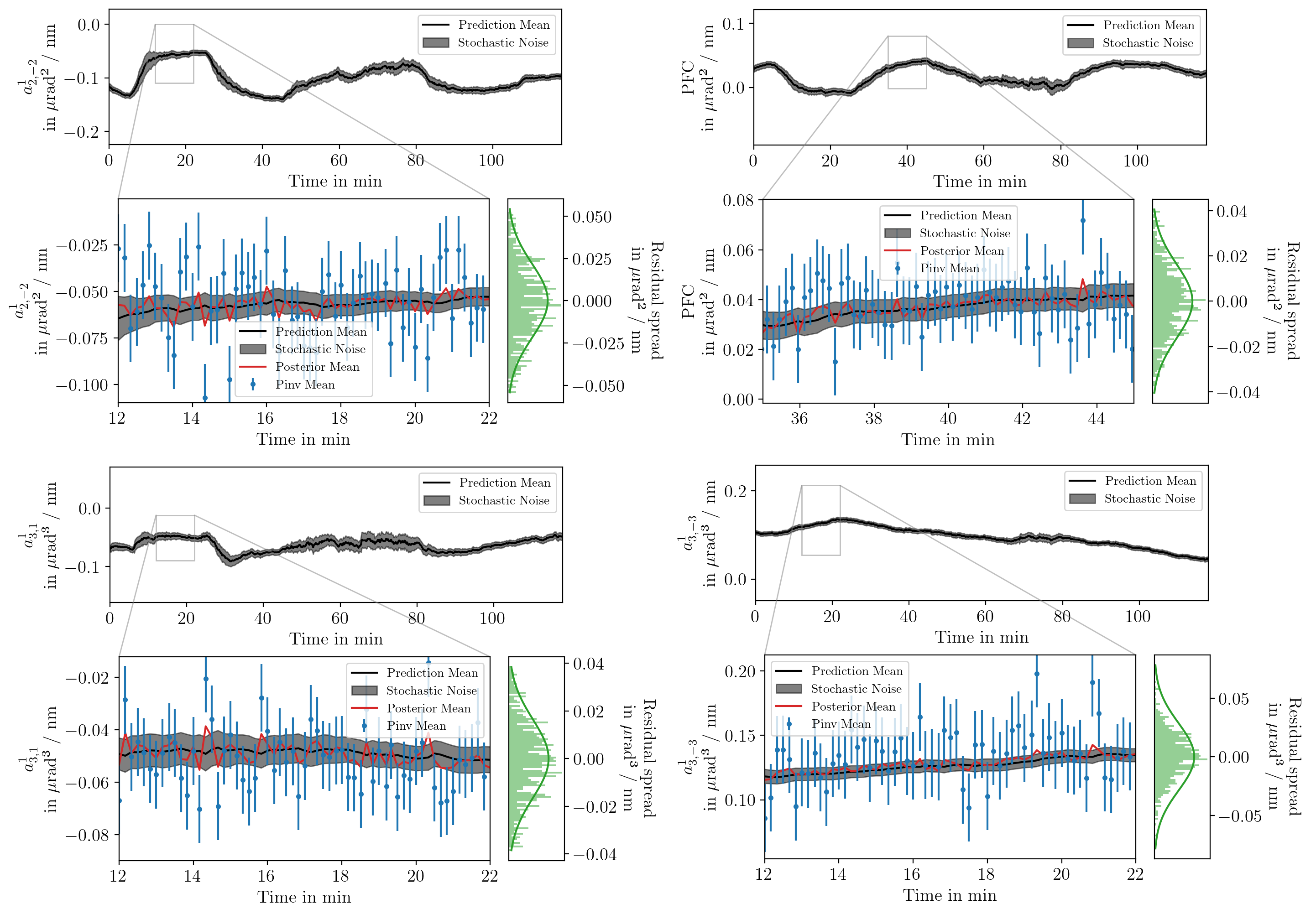}
    \caption{The evolution of the linear Astigmatism (top left), Pulse Front Curvature (PFT)/linear defocus (top right), linear trefoil (bottom left) and linear vertical coma (bottom right) prediction values together with the intrinsic stochasticity. The zoomed sections add the posterior mean as well as the pseudo-inverse (Pinv) or least-squares mean. The histograms show the spread of the residuals as well as the equivalent Gaussian distribution obtained from our estimate in Eq. 11. Respective uncertainty improvements are \SI{48}{\%}, \SI{50}{\%}, \SI{60}{\%} and \SI{47}{\%}.}
    \label{fig:linear_order_plot}
    
\end{figure*}

\end{document}